\documentclass[useAMS,usenatbib]{mn2e}
\usepackage[dvips]{epsfig}
\usepackage{amsmath}        
\usepackage{amssymb}

\newcommand{\degree}{$^{\circ}$}

\title[VLBI observations of MAXI~J1659$-$152]{VLBI observations of the shortest orbital period black hole binary, MAXI~J1659$-$152}
\author[Z. Paragi et al.]{Z.~Paragi$^{1}$\thanks{E-mail: zparagi@jive.nl}, A.~J.~van~der~Horst$^{2}$, T.~Belloni$^{3}$, 
J.~C.~A.~Miller-Jones$^{4}$, 
\newauthor 
J.~Linford$^{5}$, G.~Taylor$^{5}$, J.~Yang$^{1}$, M.~A.~Garrett$^{6,7}$, J.~Granot$^{8}$, 
\newauthor
C.~Kouveliotou$^{9}$, E.~Kuulkers$^{10}$, R.~A.~M.~J.~Wijers$^{11}$ \\
$^{1}$Joint Institute for VLBI in Europe, Postbus 2, 7990 AA Dwingeloo, Netherlands \\
$^{2}$Astronomical Institute ``Anton Pannekoek", University of Amsterdam, Postbus 94249, 1090 GE Amsterdam, Netherlands \\
$^{3}$INAF -- Osservatorio Astronomico di Brera, Via E. Bianchi 46, I--23807 Merate, Italy \\
$^{4}$International Centre for Radio Astronomy Research - Curtin University, GPO Box U1987, Perth, WA 6845, Australia \\
$^{5}$Department of Physics and Astronomy, University of New Mexico, MSC07 4220, Albuquerque, NM 87131-0001, USA \\
$^{6}$Netherlands Institute for Radio Astronomy (ASTRON), Postbus 2, 7990 AA Dwingeloo, The Netherlands \\
$^{7}$Leiden Observatory, University of Leiden, P.B. 9513, Leiden 2300 RA, the Netherlands \\
$^{8}$Departement of Natural Sciences, The Open University of Israel, P.O. Box 808, Ra'anana 43537, Israel \\
$^{9}$Space Science Office, NASA/Marshall Space Flight Center, Huntsville, AL 38512, USA \\
$^{10}$European Space Astronomy Centre (ESA/ESAC), Science Operations Department, 28691 Villanueva de la Ca\~nada (Madrid), Spain \\
$^{11}$Astronomical Institute, University of Amsterdam, Science Park 904, 1098 XH Amsterdam, The Netherlands \\
}

\begin{document}

\date{Accepted 201x Month Day. Received 2012 Month Day; in original form 2012 Month Day}

\pagerange{\pageref{firstpage}--\pageref{lastpage}} \pubyear{201x}

\maketitle

\label{firstpage}

\begin{abstract}
The X-ray transient MAXI~J1659$-$152 was discovered by Swift/BAT and it
was initially identified as a GRB. Soon its Galactic 
origin and binary nature were established. There exists a wealth of 
multi-wavelength monitoring data for this source, providing a great 
coverage of the full X-ray transition in this candidate black hole binary 
system. We obtained two epochs of European VLBI Network (EVN) electronic-VLBI 
(e-VLBI) and four epochs of Very Long Baseline Array (VLBA) data 
of MAXI~J1659$-$152 which show evidence for outflow in the early phases.
The overall source properties (polarization, milliarcsecond-scale 
radio structure, flat radio spectrum) are described well with the presence 
of a compact jet in the system through the transition from the hard-intermediate 
to the soft X-ray spectral state. The apparent dependence of source size and 
the radio core position on the observed flux density (luminosity dependent 
core shift) support this interpretation as well.
We see no evidence for major discrete ejecta during the outburst.
For the source proper motion we derive 2$\sigma$ upper limits of 
115~$\mu$as/day in right ascension, 
and 37~$\mu$as/day in declination,
over a time baseline of 12 days. These correspond to velocities of 
1400~km/s and 440 km/s, respectively, assuming a source distance of 
$\sim$7~kpc.
\end{abstract}


\section{Introduction}
Galactic black hole X-ray binaries (BHXRB) are key to understanding the power 
sources of accreting compact objects and the physical processes of accretion 
and jet formation.
A great number of these systems have been identified as transient X-ray sources, 
and sometimes they are referred to as microquasars \citep{Mirabel92,Mirabel94},
owing to their ability to produce collimated, highly relativistic ejecta analogous 
to those powered by supermassive black holes in the centres of active galactic nuclei (AGN). 
There are two types of jets in BHXRBs: compact, synchrotron self-absorbed jets
that are typical of the canonical hard spectral state 
\citep[e.g.][]{Stirling01,Dhawan00}, and transient ejecta during X-ray state 
transitions from the hard state to the soft state, crossing a region in the
X-ray luminosity vs. X-ray hardness space often referred to as the ``jet-line" 
\citep{Fender04,Fender09}. 
Whether there is a universal jet-line region for
all BHXRB, what its boundaries are, and most importantly, what the underlying 
physics is, remain a matter of debate\footnote{A recent result shows that the
location of the jet line may vary from outburst to outburst in a single object
\citep{Miller-Jones12}}. 
Most notably, we point out that high 
angular resolution observations on milliarcsecond (mas) scales have rarely been 
available in the past, although there has been an increasing number of VLBI 
observations in the past 5 years, thanks in part to the more flexible, real-time
e-VLBI operations of the EVN \citep[see e.g.][]{Tudose07}. Using the e-VLBI
technique with the EVN to observe transients has the advantage of the quick 
turnaround time, providing more efficient ways to schedule follow-up EVN 
and/or VLBA observations, as previously demonstrated for various types of transients 
\citep{Miller-Jones10, Paragi10a, Moldon11, Yang11}. The rapid feedback
is important for a better understanding of the evolution of the mas-scale source 
properties early on during the outburst event (especially for transients 
with no firm identification at their discovery). It also gives the opportunity,
as we demonstrate in this paper, to test observing strategies and look 
for closer VLBI calibrators for targets with less favourable locations, such as 
low-declination ($<-10$\degree) sources for the EVN. The focus of this paper
is a new BHXRB, which benefited from both.

On 2010 September 25 Swift/BAT discovered a new transient, which was initially
identified as a gamma-ray burst and was designated as  GRB 100925A \citep{Mangano10}.
The MAXI/GSC team detected a hard X-ray transient positionally coincident with
this GRB candidate, and catalogued it as MAXI~J1659$-$152 \citep{Negoro10}. 
Soon it became clear from optical spectroscopy by the ESO/VLT X-shooter
that the transient showed broad emission and absorption lines at zero redshift,
indicating that the source was likely a Galactic X-ray binary \citep{deUgartePostigo10}. 
The source followed an evolutionary track on the X-ray hardness-intensity 
diagram (HID) and also had variability/timing properties typical of BHXRBs, 
in particular type-B and type-C quasi-periodic oscillations (QPO) \citep{Kalamkar11}.
The X-ray lightcurves from Swift, XMM-Newton and RXTE showed irregularly
shaped dips that recurred with a period of $\sim$2.4~hours, making the source
the shortest orbital period Galactic black hole binary candidate known 
\citep{Kuulkers12,Kennea11}.
The two groups that reported these results also showed that the companion 
star is most likely an M5 dwarf, and that the system is likely located at 
a distance of 7~kpc (but see Sect.~4.1). In addition, using the scaling 
relation between the spectral index and the QPO frequency, the mass of the BH is 
estimated to be $20\pm3$~M$_{\odot}$ \citep{Shaposhnikov11}. If this estimate is 
correct, then MAXI~J1659$-$152 would be the most massive stellar black hole known 
in the Galaxy. 

The apparent high distance from the Galactic plane ($z=2.4\pm1.0$~kpc)
makes MAXI~J1659$-$152 a good candidate for being a \lq\lq runaway microquasar"
\citep{Yamaoka12,Kuulkers13}, which received a large kick velocity during the formation
of the black hole in the system. With high resolution VLBI observations
one could in principle measure the proper motion even during a single
outburst, as was demonstrated by \citet{Mirabel01} in case of XTE~J1118+480. 
The claimed mass of $20\pm3$~M$_{\odot}$ for MAXI~J1659$-$152 \citep{Shaposhnikov11}
is incompatible with the runaway microquasar scenario, therefore constraining 
the proper motion with VLBI is particularly interesting in this case.  
However, for rapidly evolving outbursts, the short timescales can make such a 
measurement quite challenging. 

Just $\sim$1.5 days after the initial trigger, 
MAXI~J1659$-$152 was detected at the 5 mJy level at 5 GHz with the 
Westerbork Synthesis Radio Telescope (WSRT) \citep{vanderHorst10a}.
This was rapidly followed by a series of EVN e-VLBI \citep{Paragi10b} 
and VLBA target of opportunity observations at 5 GHz. 
This paper summarizes the results of all the VLBI observations. We describe 
the initial e-VLBI experiment with the selection of secondary calibrators, and 
the follow-up EVN and VLBA observations in Sect.~2. Data reduction and results 
will be presented in Sect.~3. 
The compact jet model, as a framework for the interpretation of the mas-scale 
structure is presented in Sect. 4. 
In Sect.~5. we compare the mas-scale properties of MAXI~J1659$-$152 to those of other 
BHXRBs with similar evolution on an X-ray hardness-intensity diagram. 
Conclusions will be drawn in Sect.~6.

\section[]{Observational strategy and secondary calibrator selection}

The first observation of MAXI~J1659$-$152 was carried out on 2010 September 30
with the EVN at 5~GHz in realtime e-VLBI mode  with the EVN MkIV Data
Processor (``e-EVN"; experiment code RP016A). After the initial clock-searching and
setup checking, the project started at 13:00 UT and lasted 5.5 hours.
The participating telescopes were Effelsberg, Medicina, Onsala, Toru\'n, 
the phased array Westerbork Synthesis Radio Telescope (WSRT), Jodrell Bank (MkII), 
Cambridge and Hartebeesthoek. The data rate per telescope was 1024 Mbps, except 
for Medicina, Hartebeesthoek (both 896 Mbps) and Cambridge (128 Mbps, limited by 
the MERLIN microwave link to Jodrell Bank). We used 2-bit sampling and observed
both left and right circular polarizations.
The target was phase-referenced to the nearby calibrator J1707$-$1415 (2.2 degrees away),
selected from the VLBA Calibrator list\footnote{http://www.vlba.nrao.edu/astro/calib/index.shtml}.
The coordinates used for correlation of the phase reference source were $\alpha$(J2000)=17:07:20.390556 
($\pm0.25/\cos(\delta)$~mas) and $\delta$~(J2000)=$-$14:15:23.12829 
($\pm0.5$~mas). 

Because of the low declination of the field, the source elevations were low, 
especially for the EVN: the maximum elevation for most EVN telescopes was around 20 degrees, 
and it was below 20\degree\ at some telescopes for a significant fraction of the observing time. 
This limits the accuracy of determining the interferometer phase for the target using calibrator 
measurements. The coherence time is shorter, and systematic errors due to small deviations 
from the correlator delay model result in errors in the astrometry, as well as decreasing 
the fidelity of the final image. At the frequency of our observations, tropospheric and 
ionospheric phase errors have comparable effect on the data, however at low elevations the 
tropospheric errors will likely dominate. 
Both at the EVN and the VLBA correlators the tropospheric delays are determined from a zenith 
value which is then mapped to lower elevations. A small error in the tropospheric zenith delay, 
imperfections in the mapping function itself, or inhomogeneities in the troposphere will lead 
to a significant differential phase error between the calibrator and the target close to the horizon. 
This results in correlation amplitude losses, degraded image fidelity, and poor astrometry.
The closer the reference source, the smaller the differential phase errors between the reference 
and the target fields. 

Therefore we looked for additional radio sources closer to MAXI~J1659$-$152.
Unfortunately, this field is not covered by the VLA B-array FIRST survey \citep{Becker95}, 
which is an excellent resource to look for potential secondary calibrators for 
VLBI observations \citep{Frey08}. Instead, we selected 8 radio sources from the 
VLA D-array NVSS survey \citep{Condon98} that were located within half a degree of the target and had total 
flux densities exceeding 10~mJy. These were cross-checked against our 5~GHz WSRT data 
(from ongoing total flux density monitoring). One target, NVSS J170003.28$-$145622.9 
(hereafter J1700$-$1456) had a nearly flat spectrum. We included this source in the EVN
observing schedule to verify its compact structure on milliarcsecond scales.
The phase referencing cycle was 90s on J1707$-$1415 and 150s on MAXI~J1659$-$152. 
Approximately every 16 minutes we also observed the secondary calibrator J1700$-$1456.
Additional calibrators 
OQ208, J1724-1443, J1751+0939 and J1310+3220 were observed for a short time  
to check amplitude calibration of the EVN; 3C286 was included for calibrating 
the WSRT synthesis array data recorded parallel to the VLBI observations.
After the successful detection of both the transient and J1700$-$1456 (see Sect.~3.) 
we carried out further VLBI observations, all at 5~GHz. However, in the case 
of J1700$-$1456 we quickly updated the coordinates because they were found to be in error
by about one arcsecond, which is a significant fraction of the WSRT phased-array beam at 5~GHz. 
The second epoch was observed with the VLBA (BV070A) on October 2.
The VLBA data were recorded at the telescopes at a recording rate of 512 Mbps, in dual-polarization mode using 
2-bit sampling, and were correlated with DiFX \citep{Deller11}.
The third epoch was observed with the e-EVN on October 4 
(RP016B), with a similar setup to the first experiment but adding
the 40m Yebes telescope to the array. Because MAXI~J1659$-$152 was still detectable
with very long baselines \citep{Paragi10b}, additional VLBA observations were organized
(BV070B-D) on 2010 October 6, 14 and 19, until the transient faded 
below the detection threshold of our WSRT monitoring. The time between observing 
epochs was set based on the initial e-EVN results and the X-ray spectral
and timing properties as observed by RXTE.

\begin{figure}
\centering
\includegraphics[width=8cm,angle=0,bb=33 130 582 683,clip]{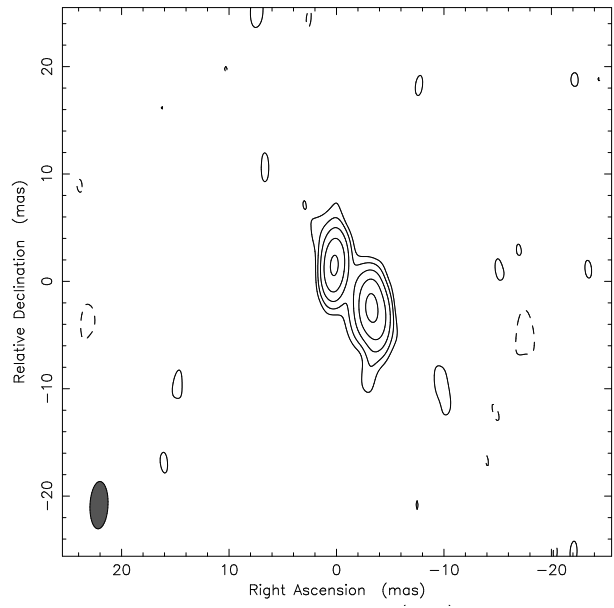}
\caption{Naturally weighted 5 GHz VLBA map of J1700$-$1456 on 2010 October 2. This source was found to be compact 
in the first e-EVN observations and was used as a secondary phase-reference calibrator at the
following epochs. The higher resolution VLBA observations show a double structure on mas scales,
but the signal-to-noise ratio was still sufficient for additional phase calibration
after the residual fringe delay, rate and phase solutions were derived using the primary 
reference source J1707$-$1415, which lies further away from the target. The peak brightness is
4.5 mJy/beam. The contour levels are $-$1, 1, 2, 4, 8 and 16 times 230 $\mu$Jy/beam. The restoring
beam is 4.4$\times$1.7 mas, with major axis PA$=-2.6$\degree\ E of N.}\label{NVSS3}

\end{figure}

\begin{figure*}
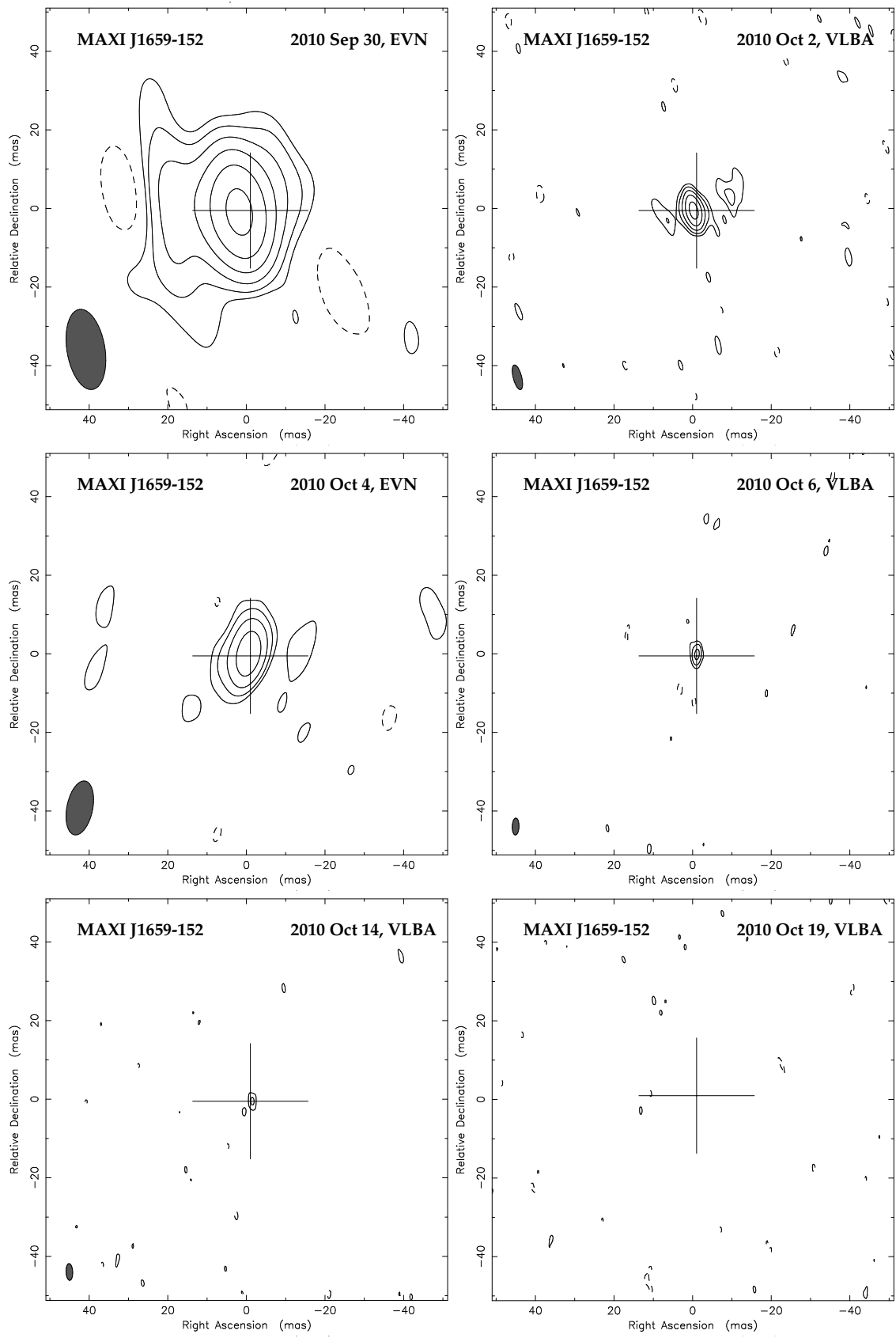

\centering
\includegraphics[width=7.4cm,angle=0,bb=33 130 582 683,clip]{MAXI_RP016A_1024.ps}
\includegraphics[width=7.4cm,angle=0,bb=33 130 582 683,clip]{MAXI_BV070A.ps}
\includegraphics[width=7.4cm,angle=0,bb=33 130 582 683,clip]{MAXI_RP016B_1024.ps}
\includegraphics[width=7.4cm,angle=0,bb=33 130 582 683,clip]{MAXI_BV070B.ps}
\includegraphics[width=7.4cm,angle=0,bb=33 130 582 683,clip]{MAXI_BV070C.ps}
\includegraphics[width=7.4cm,angle=0,bb=33 123 582 675,clip]{MAXI_BV070D.ps}

\caption{Naturally weighted 5 GHz EVN and VLBA images of MAXI~J1659-152. The coordinates of the map 
centres are $\alpha$(J2000)=16:59:01.676959, $\delta$~(J2000)=$-$15:15:28.73200. The peak brightnesses
are (in time order) 6.2, 3.9, 2.5, 0.9, 0.4 and 0.2 (non-detection) mJy/beam.
The contour levels are powers of two times the 3$\sigma$ noise level, typically
170~$\mu$Jy/beam. The cross indicates the average VLBA position. Map parameters are listed in Table~\ref{maps-models}.} \label{maps}
\end{figure*}

\begin{table*}
\caption{Map parameters for Fig.~\ref{maps}, as well as modelfit results and brightness temperatures for all observations at 5~GHz.
The columns are: observing array, observation start date [dd.mm.yyyy], Modified Julian Date of the midpoint
of the observations, peak brightness [mJy/beam], noise [mJy/beam], restoring beam major and minor axes [mas], 
position angle [degrees] (note the last epoch map is a dirty map, therefore a restoring beam is not given), 
model-fitted circular Gaussian component flux density [mJy], radius [mas], and the corresponding
(lower limit to the) brightness temperature.} \label{maps-models}
\begin{tabular}{lcrrrrrrrrr}
\hline
Array      &  Date      & MJD     &   peak   & noise        & Major & Minor   &   PA    &   $S$   & Radius& $T_{\rm b}$ \\ 
           &            &         &   mJy/b  &    mJy/b     &  mas  &  mas    &   deg.  &   mJy   & mas   &  $10^7$K    \\ 
\hline
EVN        & 30.09.2010 & 55469.7 &   6.24   &    0.052     & 20.70 & 9.67    &  9.55   &  7.31   &  5.3  &  1.4 \\ 
           &            &         &          &              &       &         &         &  1.88   & 13.0  &  0.1 \\ 
           &            &         &          &              &       &         &         &  0.46   &  2.0  &  0.6 \\ 
VLBA       & 02.10.2010 & 55472.0 &   3.92   &    0.060     & 6.45  & 2.15    &  14.6   &  6.25   &  2.4  &  6.0 \\ 
           &            &         &          &              &       &         &         &  1.58   &  6.1  &  0.2 \\ 
           &            &         &          &              &       &         &         &  0.99   &  4.7  &  0.2 \\ 
EVN        & 04.10.2010 & 55473.7 &   2.45   &    0.062     & 14.00 & 6.58    & $-$11.4 &  2.72   &  1.1  & 12.5 \\ 
VLBA       & 06.10.2010 & 55474.0 &   0.87   &    0.057     & 4.28  & 1.71    & $-$1.9  &  1.07   &  1.0  &  5.9 \\ 
VLBA       & 14.10.2010 & 55484.0 &   0.40   &    0.056     & 4.25  & 1.70    &  2.64   &  0.45   &$<0.7$ &$>4.7$\\ 
VLBA       & 19.10.2010 & 55489.0 &  $<$0.30 &    0.057     & N/A   & N/A     &  N/A    & $<$0.30 &  N/A  &  N/A \\ 
\hline
\end{tabular}
\end{table*}

\section{Data reduction and results}

The EVN and the VLBA data were reduced according to the standard procedures outlined
in the EVN Data Analysis Guide\footnote{http://www.evlbi.org/user\_guide/guide/userguide.html}
and the AIPS Cookbook\footnote{http://www.aips.nrao.edu/CookHTML/CookBook.html}, respectively.
After the initial amplitude calibration, parallactic angle correction (EVN) and Earth Orientation
Parameter correction (VLBA), the data were fringe-fitted in AIPS. 
At the first three epochs, when MAXI~J1659$-$152 was still brighter than 1~mJy, we also looked for 
polarized emission in the VLBI data. We performed cross-hand fringe-fitting on a bright
calibrator to remove the inter-channel R--L delay and phase offsets at the reference antenna.
For the EVN data we used OQ208, whereas for the first epoch VLBA dataset we used NRAO530. 
We then imaged and self-calibrated OQ208 in Difmap \citep{Shepherd94}, and the self-calibrated 
OQ208 data were used to determine the antenna polarization leakage terms with LPCAL in AIPS
(for both the EVN and VLBA data). 
The phases of both the science target and the secondary calibrator were phase-referenced to 
J1707$-$1415. The secondary calibrator, J1700$-$1456, was imaged and self-calibrated in Difmap. 
This source has resolved structure on a scale of a few mas (see Fig.~\ref{NVSS3}). 
In the higher resolution VLBA data, the average position of the maximum brightness peak in the brighter, South-Western 
component is $\alpha$(J2000)=17:00:03.333340 and 
$\delta$~(J2000)=$-$14:56:21.97584. The statistical error from the four measurements is 
170~$\mu$as in right ascension and 250~$\mu$as in declination.  Accounting for the positional 
error of the primary phase-reference source, this increases to 320~$\mu$as and 560~$\mu$as, respectively.

The image was imported back into AIPS where phase solutions were derived with CALIB using a solution 
interval of 30 minutes, and applied to the target. This step helped to improve the image fidelity of
the target, as well as the astrometric precision of the results. Note that we used the highest 
quality VLBA map to self-calibrate all epochs, except for the first epoch EVN data because the
accurate coordinates of J1700$-$1456 were not known during the correlation of the first experiment, 
therefore exact alignment of the VLBA and the first epoch EVN phase-centers is not possible below the
$\sim$mas level.  
Choosing a 30 minute solution interval ensured that we corrected for a smoothly changing residual phase
resulting from the imperfect troposphere model, rather than short timescale phase fluctuations. 
In the case of the EVN, we did not solve for Cambridge and Hartebeesthoek because they had low 
signal-to-noise ratios on J1700$-$1456. In addition, at the first epoch the secondary calibrator 
phase errors were higher than expected for the WSRT, likely because the a priori position had an error 
of about an arcsecond, which is a significant fraction of the phased-array WSRT beam. In that case 
the WSRT phases (only) were self-calibrated on the target, which had a comparable 
total flux density to the secondary calibrator at the first epoch. This additional phase 
self-calibration, using a 30 minute solution interval, did not significantly affect the final result. 
The final images of the target were made in Difmap, with no further self-calibration at the other
epochs.

In the following discussion we utilize the fact that MAXI J1659-152 had 
a flat spectrum during the first two EVN epochs. As part of a broadband 
monitoring campaign, the source was observed with the Karl G. Jansky Very Large Array (VLA) 
at multiple frequencies on 2010 September 29 and October 1. On the former date the 
flux density was $9.88\pm0.30$~mJy at 4.9 GHz and $10.03\pm0.31$~mJy at 
8.5 GHz, while on the latter date those were $10.29\pm0.32$~mJy and 
$9.74\pm0.30$~mJy, respectively. The resulting spectral indices are 
$0.03\pm0.11$ and $-0.10\pm0.11$, consistent with an index of 0. We note 
that there are no significant detections of polarization at those 
epochs, with upper limits of a few percent on the degree of 
polarization. Further details on the VLA data analysis can be found in 
van der Horst et al. (in preparation).

\subsection{Results}

The resulting VLBI maps are shown in Fig.~\ref{maps}, and the map parameters and modelfit results are
listed in Table~\ref{maps-models}.
At the first epoch on 2010 September 30 the source is well resolved, as also clearly evidenced by the
decreasing visibility amplitude with increasing baseline length seen in the $uv$-data, 
especially on the most sensitive baselines that include either Effelsberg or Westerbork. Fitting a 
single elliptical Gaussian model component to the $uv$-data gives a total flux density of 8.7~mJy 
and a characteristic size of 8~mas (well exceeding the uniformly weighted beamsize including 
all baselines), elongated at $P\!A$$\sim$132~degrees (measured from North through East). Analyzing 
the Westerbork synthesis array data taken during the VLBI observations, the total flux density of 
MAXI~J1659$-$152 was 9.8~mJy, i.e. the single component contains most of the flux density from the target. 
To recover all of the WSRT total flux density, three model components were fitted
to the data (see Table~\ref{maps-models}). Note that the naturally weighted EVN images shown in 
Fig.~\ref{maps} do not include the baselines to Hartebeesthoek, for better reconstruction of the extended 
emission. The uniformly weighted image including Hartebeesthoek data  
suggests there is extension to the South-East on mas scales, in agreement with the modelfit results. 
This image contains a total cleaned flux density of 6.3~mJy. The naturally weighted image on the other 
hand has a total cleaned flux density of 9.6$\pm0.1$~mJy, recovering practically all of the integrated flux 
density. It shows an extension roughly to the East up to 20~mas. It is hard to judge from these data 
alone whether the apparent extension to the West-North-West is real or not, but this feature agrees 
in orientation with the major axis PA of the fitted elliptical Gaussian component. 

The higher resolution VLBA image on October 2 shows a partially resolved central component 
containing 6.3 mJy. Fitting a single elliptical Gaussian component resulted in a major axis of 
$P\!A$$\sim$115~degrees, close to the single-component fit EVN value. There is extended emission on both 
sides of the source roughly in this direction. This extended emission is only detected on the shortest 
spacings. But since it does not disappear even with further point source model phase self-calibration, 
and the overall structure is similar to the EVN result, we conclude that the extended emission is likely real. 
The source was therefore fitted with three circular Gaussian components, with a total flux density of 8.8~mJy.
By the second EVN observation, on October 4, the total flux density of MAXI~J1659$-$152 had decreased to 2.95~mJy
according to our WSRT measurement. We fitted a circular Gaussian brightness distribution model directly to 
the $uv$-data. The entire integrated flux density was recovered this way in a single component (consistent
with the WSRT value within the errors), with a radius of 1.1~mas. MAXI~J1659$-$152 was detected in two 
more VLBA observations. Although the source looks compact in the images from epochs 3 and 4, the source
was still resolved according to our modelfit results (see Table~\ref{maps-models}). At the fifth epoch the fitted
Gaussian component size was smaller than the theoretical resolution limit of our array,
therefore we consider the minimum resolvable angular size as an upper limit \citep[see][]{Lobanov05}.
MAXI~J1659$-$152 faded below the detection level of 300~$\mu$Jy/beam (5$\sigma$) on 2010 October 19. 

No polarized emission was detected at any of these epochs. The first EVN epoch, with the highest 
peak in total intensity and lowest noise in polarized intensity, provides the strongest constraint. 
The 150 $\mu$Jy/beam upper limit (5$\sigma$) in polarized intensity corresponds to a 2.5\% upper limit 
in fractional polarization. 

Using the higher resolution VLBA data, the derived average position for MAXI~J1659$-$152 is 
$\alpha$~(J2000)=16:59:01.676891 and $\delta$~(J2000)=$-$15:15:28.73237. 
The error in declination is 220~$\mu$as, consistent with the measured errors in the secondary 
reference source position. In right ascension the error is much larger, 690~$\mu$as. The peak in 
the (lower resolution) first epoch EVN map is about 2~mas East of the average position. 
This is unusual since due to calibration errors at low elevations, one expects to see larger
errors in declination than in right ascension \citep[cf.][]{Yang12}. Therefore, this is 
likely the sign of a source structural effect. The October 4 EVN position is within the 2$\sigma$ 
error. The total errors in $\alpha$ and $\delta$, including the uncertainty in the primary 
phase-reference source position, are 740~$\mu$as and 550~$\mu$as, respectively.

 \begin{figure}
 \centering
 \includegraphics[width=6cm,angle=-90,bb=81 15 590 730,clip]{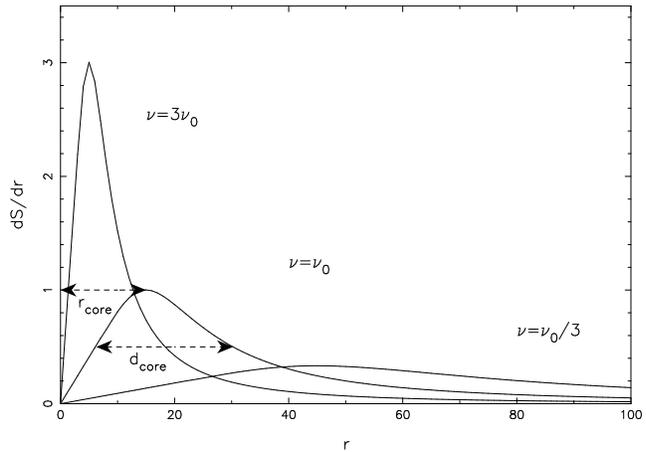}

 \caption{The jet intensity profile with distance from the central engine. Both d$S$/d$r$ and $r$ are in arbitrary units. 
$r_{\rm core}$ is the distance between the jet origin ($\sim$the compact object) and the brightest peak
of the compact jet. It demonstrates that the observed `radio cores' of microquasars and AGN do not exactly 
coincide with the location of the black hole. $d_{\rm core}$ is the FWHM size of the compact jet profile
(approximately the size of the source obtained by fitting a Gaussian profile to the $uv$-data).
The model is scaled so that the optical depth is unity at distance $r$=20, at reference frequency $\nu_{0}$. 
Jet parameters are $m$=1, $n$=2, $s$=2 ($k_{r}$=1, $k_{\tau}$=3). For an explanation of the different jet parameters, 
see the Appendix. } \label{jetprofile}

 \end{figure}

\section{Discussion}

\subsection{A compact jet in MAXI~J1659$-$152?}

The images in Fig.~\ref{maps} 
clearly show resolved or partially resolved structures at the
first three epochs, indicative of an outflow. The maximum brightness temperatures
resulting from fitting circular Gaussian components to the $uv$-data at each epoch 
range from $\sim10^7-10^8$~K (see Table~\ref{maps-models}.). These high brightness
temperatures, together with the observed flat radio spectrum (see Sect. 3),
are indicative of optically thick synchrotron radiation, which is also
consistent with the observed low fractional polarization ($<$2.5\%).
An optically thick synchrotron spectrum may originate in partially self-absorbed compact
jets (for details see Appendix). 
In fact, such partially self-absorbed compact jets have been resolved with 
VLBI in the BHXRB candidate SS433 \citep{Paragi99}, and the confirmed
BHXRB systems GRS1915+105 \citep{Dhawan00} and Cyg~X-1 \citep{Stirling01}.
Various other observational evidence points to the fact that in general, BHXRBs
produce compact jets in the canonical hard X-ray state \citep{Fender04,Fender09}.
Our highest resolution data do not show a resolved core-jet structure, 
but the images from the first two epochs clearly show extensions roughly to the East 
and to the North-West, that may be related to an outflow. The model-fitting 
results are also consistent with an elongated structure in a similar
direction. One explanation for the overall fuzzy appearance of the `core'
could be angular broadening of the images due to interstellar scattering.
Definitive evidence for scatter-broadening would be multi-frequency measurements
that showed a $\nu^{-2}$ dependence of the observed source 
size. However, the nearby secondary calibrator source J1700$-$1456 showed 
no evidence for scatter-broadening.

We have two other pieces of circumstantial evidence for supporting the self-absorbed 
jet nature for the bulk of the radio emission. Curiously, MAXI~J1659$-$152
appeared to be the most compact at the last detection on 2010 October 14,
when the total flux density was less than 1~mJy. 
At earlier epochs (and especially with the larger EVN beam)
a larger overall source size was measured because of the extended emission 
(outflow). The radius of maximum brightness temperature, and consequently the measured 
size of the compact jet region, depends on the radio luminosity \citep{BlandfordKonigl79}.
The measured size values from Table~\ref{maps-models}, 
are plotted in the left panel of Fig.~\ref{fig5} as a function of flux density. 
While the scatter is significant, there seems to be a trend of increasing angular size with 
increasing flux density, as expected for compact jets. 
The different array configurations of the EVN and VLBA observations probably affect 
the size measurements, so we cannot compare the size-flux dependence with model predictions. 
The important point here is that changes in total jet power or bulk Lorentz factor will change 
the source flux as well as apparent size: initially the source brightens and grows, but as the 
flare fades it shrinks -- this is not expected in cases where the radio emission is related 
to discrete ejecta or a shell of matter that was ejected from the system. 

A strong dependence of size with flux 
density was previously found in another microquasar, Cyg~X-3 \citep{Newell98}. In that case the
authors claimed to have observed superluminal expansion and contraction on timescales
of hours. Another similarity is that while a collimated radio jet is 
clearly present during outbursts of Cyg X-3 \cite[e.g.][]{Mioduszewski01}, the optically 
thick base morphologically does not resemble a core-jet structure.  Instead,
it looks like the `fuzzy' core of MAXI~J1659$-$152. In the case of Cyg~X-3, the
scatter-broadening effect is well established, with a scattering disk size of
$\sim$18~mas at 5~GHz \citep{Mioduszewski01}. However, the explanation of scattering
is inconsistent with the small size we measured in the last epoch.
This same argument (i.e.\ the measured source size being smaller than the 
scattering size) could also imply that \cite{Newell98} 
may not have measured actual physical sizes, thereby weakening the claim for apparent 
superluminal expansion in Cyg~X-3 \citep{Mioduszewski01}. In the case of MAXI~J1659$-$152, 
a scattering screen, if present, must be well localized at the target position because 
scatter-broadening is not observed in the nearby calibrators.

\begin{figure*}
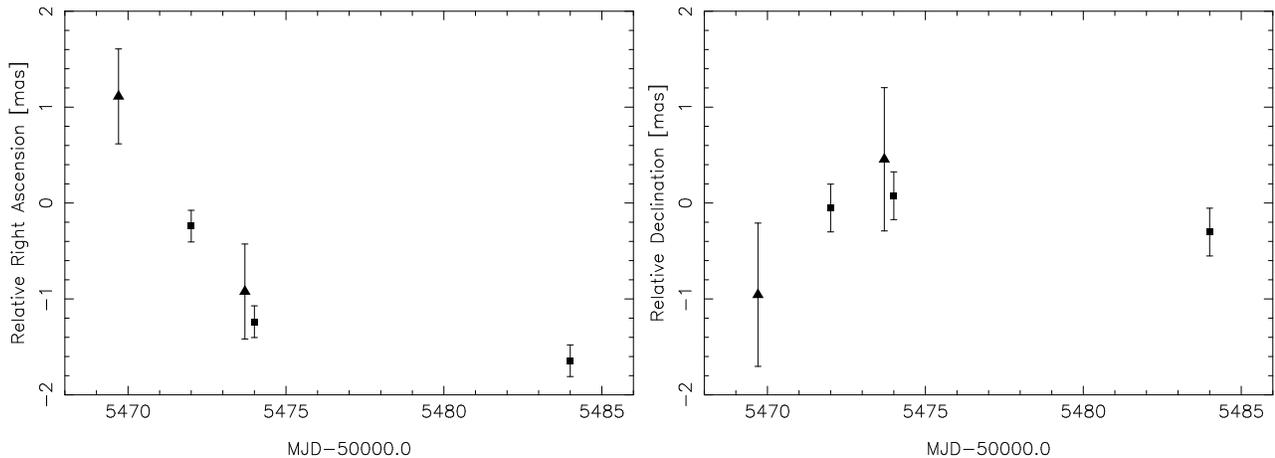

\centering
\includegraphics[width=6.0cm,angle=-90,bb=100 35 575 700,clip]{RA-MJD.ps}
\includegraphics[width=6.0cm,angle=-90,bb=100 35 575 700,clip]{DEC-MJD.ps}
\caption{Map peak position relative to the map centre in right ascension ($\alpha\cos\delta$; left panel) 
and declination ($\delta$; right panel), as measured at 5~GHz at the various epochs. 
The EVN measurements are indicated by triangles, the VLBA ones with squares.
There is a clear, smooth change in right ascension during the VLBI monitoring 
programme, most likely related to structural changes in the emission.
The increasing negative offset in right ascension shows a gradual shift of the
source peak emission to the West. There is no clear variation in declination, and the 
measurements agree within the errors. 
} \label{positions}
\end{figure*}

\begin{figure*}
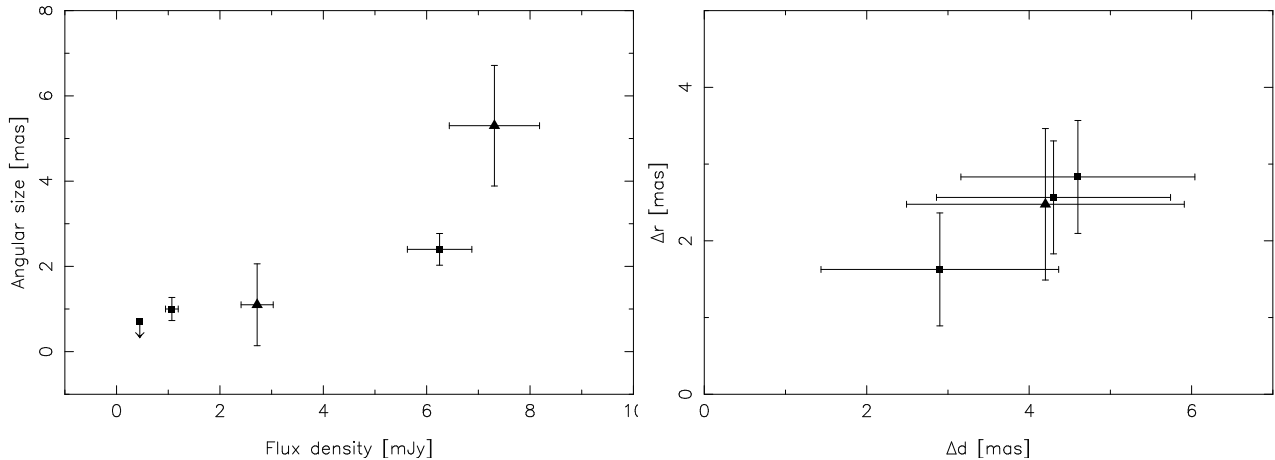

\centering
\includegraphics[width=6.0cm,angle=-90,bb=100 35 575 700,clip]{size_flux.ps}
\includegraphics[width=6.0cm,angle=-90,bb=100 35 575 700,clip]{size_coreshift.ps}
\caption{The left panel shows the source size of MAXI~J1659-152 (circular Gaussian component fit;
we used the dominant core components at the first two epochs) as a function of its flux density 
at 5~GHz (triangles: EVN; squares: VLBA). 
In the right panel we show the observed `core' position change ($\Delta\,r$)
versus the measured source size change ($\Delta\,d$) relative to the first epoch.
From a simple scaling of the opacity in our compact jet model, it is expected that the 
position of the peak emission and the FWHM width of the compact jet scale linearly with one another. 
Therefore, by analogy with the frequency-dependent core shift (see Fig.~\ref{jetprofile}.), one
may measure a time-dependent core shift at a fixed frequency during an outburst,
and compare it to the observed change in core size. The data seem to show the expected trend.
Admittedly, the number of measurements is small, and the errors are large (triangle: EVN; squares: VLBA).
} \label{fig5}
\end{figure*}

Besides the apparent correlation between the source core flux density and its size,
we compare the dependence of the apparent core position with the variable core size, 
because in compact jets these are intimately related. In Fig.~\ref{jetprofile} we show the 
intensity profile for a simple compact jet model (see Appendix). 
The observed radio core position as well as its apparent size change with 
observing frequency. This is because at different frequencies the optical depth 
is different; an effect known as the frequency dependent core-shift in AGN 
\citep{Lobanov98}. The true position of the black hole is not easy to 
determine with high accuracy. \citet{Hada11} used multi-frequency observations
to pinpoint the location of the jet base in M87, and they concluded that the supermassive
black hole is within 14--23 Schwarzschild radii of the 43~GHz radio core 
(a projected angular separation of 41~$\mu$as).  
The only similar measurement for Galactic microquasars was made in SS433. 
\citet{Paragi99} used the well-established kinematic model of the ballistically
moving, optically thin jet components to locate the binary system in between the
approaching and receding sides of the optically thick radio core components.
The separation between the binary system and the observed radio core position
decreased with increasing frequency, i.e. they observed the frequency dependent 
core-shift in a Galactic microquasar.  

Because we observed MAXI~J1659$-$152 at a single frequency, we cannot probe
the frequency dependent core shift. However, any change of optical depth
during an outburst (due to e.g. an increase in relativistic particle number density and/or 
an increase in the magnetic field strength) will result in a similar effect when we compare 
core positions and sizes at various epochs at a single frequency. 
This will result in a larger separation of the observed peak radio emission and the 
compact object, and at the same time a larger observed size for the compact jet.
This effect can be probed with VLBI at a single frequency, even if, as mentioned above, the core shift,
$r_{\rm core}$, cannot be measured in most cases. By simply scaling our model with 
the initial optical depth, we find that $r_{\rm core}$ is roughly proportional to 
the FWHM of the jet profile, $d_{\rm core}$, in which case the difference of peak positions $\Delta r$ will be 
proportional to the difference of the measured core sizes $\Delta d$ between the 
various epochs. In Fig.~\ref{positions} we plot the relative peak positions of MAXI~J1659$-$152. 
Note the apparent positions in right ascension continuously drift from roughly East 
towards the West at the various epochs, indicating either proper motion or varying 
source structure. Because the first epoch map clearly shows an extension to the 
East, it is plausible that we see a shrinking compact jet as the source flux density 
decreases. In the right panel of Fig.~\ref{fig5} we plot the difference in core position 
versus difference in core size, with respect to the first epoch. 
The data show a smooth change in source size along with a smooth change
in core (more precisely, compact jet peak) position in a way that is compatible with the 
expected relation between $r_{\rm core}$ and $d_{\rm core}$ for compact jets.
Monitoring VLBI observations of similar transients at multiple frequencies can 
potentially confirm these effects: the core size and position change with luminosity (
luminosity dependent core-shift), as well as with observing frequency (frequency 
dependent core shift, shown in Fig.~\ref{jetprofile}).  

\begin{figure*}
\centering
\includegraphics[width=15.0cm,angle=0,bb=0 0 1344 695,clip]{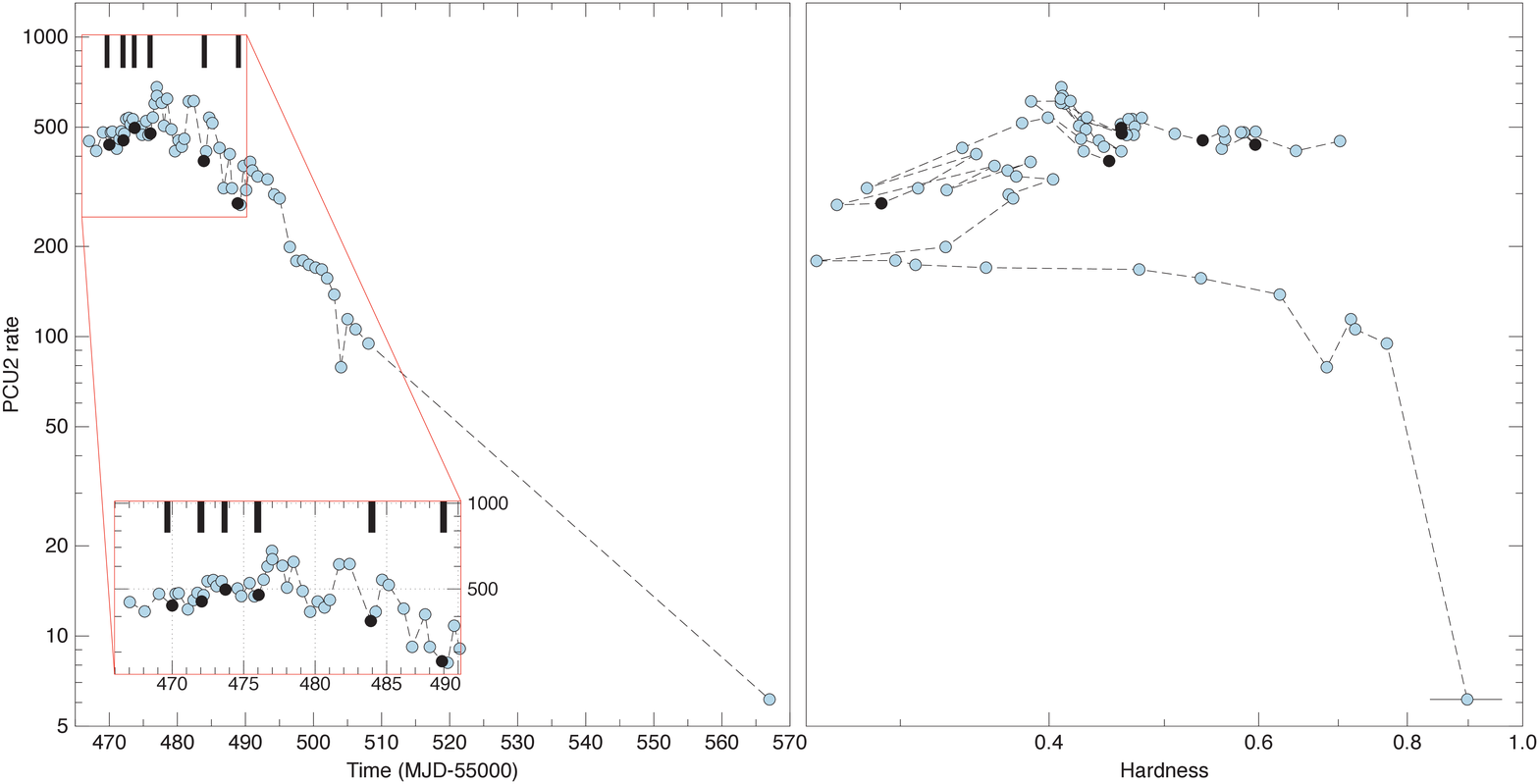}
\caption{RXTE PCU2 count rate (2--20 keV, left) and hardness-intensity diagram
of MAXI J1659$-$152 during its outburst in 2010. The hardness is
defined as the ratio of count rates measured in the 6--15 keV and
the 2--6 keV energy ranges. The black bars indicate the times of the
VLBI observations while the black circles show the RXTE 
measurements closest in time to the VLBI observations.}\label{RXTE}

\end{figure*}

On the other hand, one may interpret the positional change as source proper motion. Here we 
take a conservative approach. Because the first epoch was clearly affected by source structure
and the time baseline is very short, we give a 2$\sigma$ upper limit of 1380~$\mu$as in
right ascension and 440~$\mu$as in declination; in twelve days (epochs 2--5), these
correspond to proper motion upper limits of $\mu_{\alpha}<115$~$\mu$as/day and
$\mu_{\delta}<37$~$\mu$as/day\footnote{We note that in order to minimize the structural effects 
in right ascension, one could use the last two VLBA detections to constrain the proper
motion in right ascension, when the source was most compact. Although the time baseline is 
somewhat shorter (9 days), the measured statistical errors in right ascension for the secondary 
calibrator are also smaller. Using these we would arrive at a very similar proper motion
constraint in right ascension as in declination.}. 
To convert these to velocities, we have to know the source distance, 
which is not strongly constrained, as is typically the case for X-ray transients. 
Based on the relation between optical outburst amplitude and the orbital period
in similar systems \citep{Shahbaz&Kuulkers98}, \citet{Kuulkers12} estimated a distance 
of 7$\pm3$~kpc for MAXI~J1659-152.
As pointed out by \citet{Miller-Jones11}, this does not take into account the effect of
orbital inclination to the line of sight; at high inclinations the observed brightness of the
disc would be underestimated, and consequently, the derived distance would be an overestimate.
In a subsequent paper, \citet{Kuulkers13} derived an updated distance of 8.6$\pm$3.7~kpc, 
using the original amplitude-period relation.  
Correcting for the inclination angle of the orbit, the distance would
be lowered to 7.1$\pm3.0$~kpc at $i=65^\circ$, and 4.0$\pm1.7$~kpc for $i=80^\circ$.
The corresponding heights above the Galactic plane are $z=2.4\pm1.1$~kpc, $z=2.0\pm0.9$~kpc and
$z=1.1\pm0.5$~kpc, respectively \citep{Kuulkers13}.
Our VLBI images do not show a highly beamed source structure, with a very compact core and 
one-sided ejection, and they do not show symmetric double-sided ejecta in the plane of the sky
either. Assuming that the jets are perpendicular to the accretion disk, this would mean that 
the orbit of the binary is likely moderately inclined to the line of sight, preferring the
$\sim$7~kpc value. Finally, by carefully taking into account various arguments, \citet{Jonker12}
estimated a source distance of $6\pm2$~kpc, although this error bar may be underestimated given 
the number of assumptions involved.

Assuming a distance of 7~kpc, our proper motion constraints correspond to upper limits on the 
space velocity of $\lesssim$1400~km\,s$^{-1}$ in right ascension and $\lesssim$440~km\,s$^{-1}$ 
in declination. In the high inclination angle case (corresponding to a distance of 4~kpc),
these numbers would change to $\sim$800~km\,s$^{-1}$ and $\sim$250~km\,s$^{-1}$, respectively. 
These values are not particularly constraining, since there are no measured black hole kicks as large as 
1400 km\,s$^{-1}$. \citet{Yamaoka12} argued that MAXI~J1659$-$152 is a runaway microquasar, just
like XTE~J1118+480, which has a velocity of 145~km\,s$^{-1}$ with respect to the Local Standard of Rest
\citep{Mirabel01}. \citet{Kuulkers13} concluded that 
this was a likely scenario for MAXI~J1659$-$152 because shorter period (less massive) systems should
have the highest runaway velocities; this scenario would also explain the large distance from the
Galactic plane. Our twelve day time baseline was too short to significantly 
constrain the velocity. With the accurate coordinates measured during this outburst, future 
VLBI observations in the next activity cycle will give an accurate space velocity measurement 
for MAXI~J1659$-$152. For future X-ray transients with a more prolonged flaring activity, it will be possible 
to use this method to probe the runaway microquasar scenario if the transient is within a few kpc. 
Further improvement on the positional accuracy can be obtained if one 
can find a secondary calibrator closer to the target, especially one within the primary beam of 
the telescopes. But as we have seen in here, varying source structure will likely be a
limiting factor in most cases.



\subsection{X-ray properties and radio structure}

The observed radio structure of MAXI~J1659$-$152 is consistent with a compact jet,
which is typical for the canonical hard state of BHXRBs. We will now consider the 
X-ray properties of the transient. The RXTE lightcurve and HID (6$-$15 keV/2$-$6 keV) 
are shown in Fig.~\ref{RXTE}; black points show measurements closest to the VLBI 
epochs. The source already left the
canonical hard state and was in the hard intermediate state (HIMS) by the time of 
the first EVN epoch. The presence of a compact jet in the HIMS state is not unprecedented; 
Cyg X-1 is known to show a compact jet in this state. During the evolution to the soft 
intermediate state (SIMS), radio quenching was reported from MAXI~J1659$-$152 on October 8 
\citep{vanderHorst10b} and type-B QPOs appeared on October 12 \citep{Munoz-Darias11},
which are usually followed by relativistic jet ejections. Note that radio quenching
and subsequent bright, discrete ejecta are typical of BHXRBs during their transition
to the soft state. However, while our measurements confirm the radio quenching, we do not 
see discrete ejecta in MAXI~J1659$-$152. Instead, the source simply faded 
below our detection limit. 
Because of the relatively rapid cadence ($\sim$2 days) of our observations 
during the radio bright state of the source, and because no other flaring events
were reported from total flux density monitoring campaigns, it is quite unlikely 
that we missed an ejection event. We note also that our final VLBA observation
was taken when the source was in its softest X-ray state (see Fig.~\ref{RXTE}), and by that
time the RXTE count rate was well into its decay phase, therefore we do not
expect significant radio flaring after this time. The lack of ejecta may indicate that  
MAXI~J1659$-$152 underwent a failed transition.
However, as was shown by \citet{Munoz-Darias11}, the X-ray state transition was
fully completed, according to the well established criteria: the total rms variability
went down to almost 1\%, unlike H1743$-$322. It is true however
that the soft state in MAXI~J1659$-$152 was very short-lived and appeared
significantly harder than that in most BHXRBs. \citet{Munoz-Darias11} proposed that this
may be due to the high inclination of the system, in which case we see less of the soft
disc emission while the emission from the spherical, non-thermal corona should be
largely inclination independent. A similar lack of bright, transient ejecta was reported 
for the transition of Cyg~X-1 to the soft state in 2010 June \citep{Rushton12}. However, Cyg~X-1 
is also peculiar in that it has a non-canonical soft state (significantly higher 
fractional rms variability than typical transient BHXRB outbursts). VLBI observations
of many more transient events are necessary to draw general conclusions about the 
formation of transient ejecta, whether they appear in all soft-state transitions,
and where exactly the `jet-line' is located in the hardness-intensity diagram
\citep[which may vary from system to system, and even within a given system; ][]{Miller-Jones12}.

\citet{Jonker12} discuss the well-known radio--X-ray correlation in BHXRBs 
\citep[e.g.][]{Corbel03,Gallo03,Jonker04}. During the canonical hard state, the observed 
correlation is $L_{\rm R}\propto L_{\rm X}^{0.6}$; more recently it has been shown that 
a number of systems follow a much steeper correlation with  
$L_{\rm R}\propto L_{\rm X}^{1.4}$ \citep{Coriat11,Gallo12}. 
At the beginning of its outburst, while the luminosity was in excess of 
$10^{36}$~erg\,s$^{-1}$, \citet{Jonker12} found MAXI J1659-152 to 
follow the steeper correlation, but to be significantly more radio 
luminous at a given X-ray luminosity than the rest of the sources on the 
steeper correlation track. To explain this discrepancy, they proposed 
that the source either falls on the outlier branch, but with suppressed 
X-ray luminosity owing to the high inclination, or that the radio 
emission was possibly dominated by optically thin ejecta at early times, 
increasing the average radio luminosity. In contrast, in figure 9 of 
\citet{Corbel13}, MAXI J1659-152 does not appear to be significantly 
radio overluminous for the steeper correlation track. A re-examination 
of the original data of \citet{Jonker12} revealed a discrepancy in 
the assumed distance of H1743-322, the single source that best defines 
the steeper correlation track \citep{Coriat11}. Assuming instead the 
accepted distance of 8~kpc for H1743-322, MAXI J1659-152 would no longer 
appear significantly radio overluminous (P. Jonker, priv. comm.).  
However, this does not address an additional motivation for the 
possibility of optically-thin radio flares namely the reported detection 
of a high level of linear polarisation from a preliminary analysis of 
WSRT data \citep{vanderHorst10a}.  Reanalysis of these WSRT data 
however showed no significant linear polarization (see Sect. 3), in 
agreement with our VLBI results, which show no evidence for optically 
thin ejecta. The small observed physical size, the flat spectrum, the 
lack of significant proper motion and the low linear polarization 
suggest that the radio emission was dominated by a compact jet.

\section{Conclusions}

We presented a series of VLBI (realtime e-VLBI at the EVN plus VLBA) monitoring results
for the BHXRB candidate MAXI~J1659$-$152. The VLBI maps show evidence for an outflow 
on a scale of 10-mas. The compact jet scenario fits the data well: the low polarization, 
flat spectrum as well as the observed source size and luminosity dependent core shift 
all support this interpretation. During the full phase transition from the HIMS through 
the SIMS to the soft state, we do not see evidence for the luminous, discrete ejecta 
that are typical for BHXRBs in this state, possibly indicating that the outflow rate 
was not sufficient to form powerful shocks, as seen in the case of Cyg~X-1. 
As \citet{Yamaoka12} and \citet{Kuulkers13} pointed out, MAXI~J1659$-$152 could 
potentially be a runaway microquasar. Further VLBI observations 
during another outburst will be necessary to measure the proper motion of the system 
with high accuracy. With the method presented here, it should be
possible to provide useful constraints on black hole transient proper motions even during
a single outburst if the outburst lasts long enough.  The runaway microquasar hypothesis
could then be tested.

\section*{Acknowledgments}
We thank the anonymous referee for the constructive comments, which helped to 
improve our paper significantly. ZP thanks to Andrei Lobanov for discussions 
about the compact jet model. e-VLBI research infrastructure in Europe is 
supported by the European Union's Seventh Framework Programme (FP7/2007-2013) 
under grant agreement RI-261525 NEXPReS. The EVN is a joint facility of European, 
Chinese, South African and other radio astronomy institutes funded by 
their national research councils. The WSRT is operated by ASTRON 
(Netherlands Institute for Radio Astronomy) with support from the Netherlands 
foundation for Scientific Research. The National Radio Astronomy Observatory is a 
facility of the National Science Foundation operated under cooperative agreement 
by Associated Universities, Inc. This work made use of the Swinburne University of 
Technology software correlator, developed as part of the Australian Major National 
Research Facilities Programme and operated under license. The research leading to 
these results has received funding from the European Community's Seventh Framework
Programme (FP7/2007-2013) under grant agreement ITN 215212.

\appendix
\section{Compact jet model}

A conical jet model to describe the intensity profile of the radio jets in
the galactic XRB SS433 was developed by  
\citet{HjellmingJohnston88}, hereafter HJ88. Here we follow HJ88, but
instead of the large-scale jet we will concentrate on the properties
of the optically thick inner core, also taking into account more general
considerations for the variation of magnetic field with distance, as done for
case of AGN jet models (see below). Let us assume a conical and freely 
expanding jet with radius proportional to the distance $r$ 
to the central engine. Synchrotron radiation is emitted by electrons 
moving in a tangled magnetic field with average strength $B$. We assume 
that the energy spectrum of the radiating electrons -- filling the entire 
volume of the jet -- has a power-law form,
$N(\gamma_{e})\sim \gamma_{e}^{-s}$, where $s$ is the energy spectral 
index, which is generally thought to be $\sim$2 ($1.0\la s\la 3.0$ 
will be considered in the present paper). The optically thin spectrum 
of the source will have a spectral index 
$\alpha =(1-s)/2$ ($S\propto \nu^{\alpha}$; e.g. \citet{RybickiLightman79}).
The magnetic field and the particle density are assumed to depend
on $r$ as $B\propto r^{-m}$ and $N\propto r^{-n}$; 
for a conical geometry, $m$ can vary between 1 and 2 \citep{BlandfordRees74}. 
If $m=1$ and $n=2$, the energy density of the magnetic field and the kinetic 
energy density of the electrons are in equipartition at any $r$ along the 
flow \citep{BlandfordKonigl79}. All quantities determined below are expressed 
in the emitting frame of the fluid. 

The optical depth to synchrotron radiation through the jet axis 
can be written as \citep[cf.][]{Lobanov98}
\begin{equation}
\tau (r)=\tau_{0} (r/r_{0})^{-k_{r} (\epsilon+1)} (\nu /\nu_{0})^{-(\epsilon +1)},
\label{tau}
\end{equation}
where $\epsilon=(s +2)/2$ and $k_{r}=(\epsilon m+n-1)/(\epsilon +1)$.
For the equipartition values of $m=1$ and $n=2$, $k_{r}=1$ independent
of the energy spectral index. In this case the optical depth scales 
as $r^{-(\epsilon +1)}$ and the location of the radio core
is inversely proportional to the observing frequency \citep{BlandfordKonigl79}.
The actual value of the constant $\tau_{0}$ depends on the selection 
of $r_{0}$ and $\theta$, the jet inclination to the line of sight 
(assumed to be constant). 

Following HJ88, the contribution to the 
flux density of a portion of the jet of length $dr$ can be written  
\begin{equation}
\frac{dS}{dr} = C\sin\theta\, (\nu/\nu_{0})^{5/2}\,
(r/r_{0})^{(m+2)/2}\, (1-e^{-\tau})\, \xi_{con}(\tau),
\label{profile}
\end{equation}
where $\xi_{con}=a+b\tau -c\tau^{2} ...$ is the geometrical correction function
for conical jets and $a$, $b$, $c$ \ldots \, constants are given by 
\citet{HjellmingHan95}.
The actual value of the constant $C$ is determined by the opening angle of the flow,
the magnetic field strength and number density of electrons, etc., at a reference 
distance $r_{0}$.
Note that Eq. (\ref{profile}) and Eq. (\ref{tau}) are more general than that 
given originally by HJ88, since $m$ and $n$ are not fixed. These values, 
together with the energy spectral index $s$  -- through $k_{r}$ -- 
determine the shape of the jet intensity profile (see Fig.~\ref{jetprofile}).  

The jet intensity first rises sharply with $r$ and reaches its maximum at
$r=r_{\rm core}$ (at a given frequency).
Within a few $r_{\rm core}$ its flux density will decrease significantly. 
One can also determine the brightness variation in the core with
frequency. The optically thick limit of the integrated flux density 
from Eq.~(\ref{profile}) gives \citep{Konigl81}
\begin{equation}
\alpha_{\rm core}
 = 5/2 - (m+4)/(2 k_{r})
\label{Sratio}
\end{equation}
While the peak of the jet intensity profile is roughly proportional
to the frequency, the integrated spectrum of the core remains flat or
slightly inverted, as typically observed in radio-loud AGN. The spectral index
can vary between $-0.5\la \alpha \la 1.5$, with the equipartition
value of $\alpha=0.0$ .

In our model $k_{r}$ is assumed to be constant throughout
the region of interest (in the core at a given frequency). If $k_{r}$
changes smoothly with decreasing $r$ (consequently, at increasing $\nu$),
then Eq.~(\ref{profile}) is approximately valid only in a frequency range
$\Delta\nu=\nu_{2}-\nu_{1}$, for which $k_{r}$ is determined.
In the inner jets (at $\nu\gg 1$~GHz) however,  the geometry 
of the flow may be different from conical due to external pressure gradients
\citep[e.g.][]{GeorganopoulosMarscher96} 
and $k_{r}$ may change very fast. Clearly, our simple model does not describe
this scenario well. External pressure gradients may be caused for example by 
a dense, optically thick stellar wind. As most of the jet environment is 
expected to be ionized, this results in additional free-free absorption, 
and the jet intensity profile will depend also on the distribution of the 
absorbing medium.

\bsp

\label{lastpage}

\end{document}